\documentclass[ amsfonts, amssymb, amsmath,twocolumn,linenumbers]{revtex4-1}
\usepackage{amsmath}
\usepackage{amssymb}
\usepackage{graphicx}
\usepackage{color}
\usepackage[T1]{fontenc}
\usepackage{lineno}
\begin{document}
\nolinenumbers
\title{Modelling Covid-19 epidemic in Mexico, Finland and Iceland}
\author{Rafael A. Barrio$^{1}$}
\author{Kimmo K. Kaski$^{2,3}$}
\author{Gu\dj mundur G. Haraldsson$^{4}$}
\author{Thor Aspelund$^{5,6}$}
\author{Tzipe Govezensky$^{7}$}
\affiliation{$^1$Instituto de F{\'i}sica, Universidad Nacional Aut{\'o}noma de  M{\'e}xico, CP 01000 CDMX, Mexico}
\affiliation{$^2$Department of Computer Science, Aalto University School of Science, FI-00076 AALTO, Finland}
\affiliation{$^3$The Alan Turing Institute, 96 Euston Rd, Kings Cross, London NW1 2DB, UK}
\affiliation{$^4$ Science Institute, University of Iceland, Dunhaga 3, 107 Reykjavik, Iceland}
\affiliation{$^5$ Centre for Public Health Sciences, University of Iceland, Reykjavik, Iceland.}
\affiliation{$^6$The Icelandic Heart Association, Iceland.}
\affiliation{$^7$Instituto de Investigaciones Biom\'edicas, Universidad Nacional Aut\'onoma de M\'exico, Apartado Postal 70228, Ciudad Universitaria, CP 04510 CDMX, Mexico}

\date{\textrm{\today}}

\begin{abstract}
Over the past two decades there has been a number of  global outbreaks of viral diseases. This has accelerated the efforts to model and forecast the disease spreading, in order to find ways to confine the spreading regionally and between regions. Towards this we have devised a model of geographical spreading of viral infections due to human spatial mobility and adapted it to the latest Covid-19 pandemic. In this the region to be modelled is overlaid with a two-dimensional grid weighted with the population density defined cells, in each of which a compartmental SEIRS system of delay difference equations simulate the local dynamics (microdynamics) of the disease. The infections between cells are stochastic and allow for the geographical spreading of the virus over the two-dimensional space (macrodynamics). This approach allows to separate the parameters related to the biological aspects of the disease from the ones that represent the spatial contagious behaviour through different kinds of mobility of people acting as virus carriers. These provide sufficient information to trace the evolution of the pandemic in different situations. In particular we have applied this approach to three in many ways different countries, Mexico, Finland and Iceland and found that the model is capable of reproducing and predicting the stochastic global path of the pandemic. This study sheds light on how the diverse cultural and socioeconomic aspects of a country influence the evolution of the epidemics and also the efficacy of social distancing and other confinement measures.
 \end{abstract}
\maketitle

\section{Introduction}
In today's globalised world people easily travel long distances to faraway places, where they can get infected, carry and consequently spread lethal and easily mutating viruses that could subsequently lead to wide-spread epidemic or even pandemic of catastrophic proportions. Recent examples of such globally spreading viral diseases are the seasonal influenza, SARS, MERS, AH1N1 swine flu and currently COVID-19 that is turning out to be rather lethal showing growing death tolls in various countries. This in turn causes a lot of concern among general public and huge strain in national healthcare and hospital capacities including their Intensive Care Units (ICUs), but also among public authorities and governments as to what measures, restrictions, recommendations to put in place and when, to keep the society running in balance with the rest of its socioeconomic functions. 

In order to make decisions of these large scale, and far reaching issues for mitigating the effects of the disease, one would need to take into account the pathological aspects of the virus, epidemiological factors including mechanisms of virus spreading, human sociality including their social contacting, behavioural, mobility, travelling and communication patterns, demographic factors including age, gender, and regional density of population as well as concentration in built environments, especially cities and public transportation volumes, terrestrially and aerially. 

The commonly accepted measures to counter the effect of these factors are recommendations for individuals in terms of enhanced hygiene and keeping 1 to 2 meter social or in fact physical distance, or the government or local authority enforced restrictions for staying and working at home in self-quarantine, requirement to wear face mask in public places and transportation or even curfews, allowing  gatherings with only a small number of people, lock-down of countries or districts by closing schools, universities and other public institutions as well as restaurants and shops, other than food markets, limiting public local, terrestrial and aerial transportation as well as non-essential entries to the countries or regions. 

In addition, various countries have adopted comprehensive testing of even asymptomatic individuals together with tracking them with smart phone app. It is evident that countries adopt, follow, and ease these restriction and confinement measures in a number of different ways and with different timings, as many of them tend to be strongly linked to the culture, politics, economy and overall well-being of the country. Furthermore, there are collective effects due to news and social media depending on how lethal the disease is perceived to be.

Over the past two decades a lot of effort and advance have been made to get understanding of the population level complex dynamics of infectious disease spreading~\cite{and, bailey,cast}. In this various epidemiological models -  often compartmental SIRS (susceptible - infected - recovered - susceptible) type -  have proved very versatile in implementing the above described epidemically relevant factors for qualitative and quantitative or predictive insight into the cause of infection spreading~\cite{johansen, london, bootsma, viboud}. Models including geographical spread and population heterogeneity have also been developed~\cite{longini05, ferguson, Balcan, barabasi, barrio, apolloni, marguta, santermans, chowell}, and recently, many models including different aspects of pandemics have been developed, see for example~\cite{chen, chen2, yang, bekiros}.

Therefore, it is interesting to study how these social restrictions work and how efficient they are in countries with different cultures and social habits. In this work we decided to do a comparative study of three very different countries, Mexico, Finland and Iceland. They differ not only in population size and density, but also in case of pandemic in their behavioural habits, restriction and confinement policies, testing and tracking as well as how individuals follow and respond to them and to authorities. For this purpose we use a model~\cite{barrio} devised to deal with the viral swine flu pandemics that started in  Mexico in 2009. The model is different from former studies~\cite{Balcan, ferguson, longini05, santermans}, the aims of which were to mimic the development of a particular epidemic by evaluating parameters from specific data. This model enables us to study not only the actual observed data, but also to predict the temporal evolution of the disease under hypothetical and changing scenarios of social distancing and confinement measures in a given geographical region. 

\section{Model}
\subsection{SEIRS epidemiological model (Microdynamics)}

The model used in this study is an extension of the one presented earlier by Barrio et al.~\cite{barrio}, and it considers the geographical spread of the epidemics at two levels: the local dynamics and more global transmission of the disease from one place to another. For this purpose, a two-dimensional grid of a country or region is constructed in such a way that within each cell $(i,j)$ the population density $\rho(i,j)$ is considered to be homogeneous. For the local dynamics a Susceptible - Exposed - Infected - Recovered (-Susceptible) or SEIR(S) model ~\cite{jose} is used for each cell, consisting of four compartments: not infected susceptible ($X$), exposed yet not infectious ($E$), infectious ($Y$), and recovered temporary immune ($Z$), representing an instantaneous local average state of the population.

In the model the periods of latency ($\epsilon$), infectiousness ($\sigma$) and immunity ($\omega$) are assumed to be constant and dimensionless by expressing them in units of a time scale $\tau$ of one day. Here we assume that the population size does not change throughout the simulation, i.e. $N=X+E+Y+Z= constant$ and for the mortality we assume an exponential functional form with a constant rate $\mu=1/L,$ where $L$ is the life expectancy, and the birth rate is $\mu N$ with all the newborn considered susceptible. As people who have recovered do not necessarily become susceptible again, because of long lasting immunity or changes in social behaviour, the actual fraction of population that becomes susceptible ($S$) is included. 
 
Considering all these assumptions, we can now express the flow rates per day for all the variables by the following map equations on each cell $\alpha=(i,j)$:

\begin{equation}
\begin{split}
\label{rates}
&X_{t+1}(\alpha)=q\left[ X_t(\alpha)-G_t(\alpha)+Sq^{b}G_{t-1-b}(\alpha)\right] +\mu N,\\
&E_{t+1}(\alpha)=q\left[ E_t(\alpha)+G_t(\alpha)-q^{\epsilon}G_{t-1-\epsilon}(\alpha)\right],\\
&Y_{t+1}(\alpha)=q\left[ Y_t(\alpha)+q^{\epsilon}G_{t-1-\epsilon}(\alpha)-q^{a}G_{t-1-a}(\alpha)\right],\\
&Z_{t+1}(\alpha)=q\left[ Z_t(\alpha)+q^{a}G_{t-1-a}(\alpha)-q^{b}G_{t-1-b}(\alpha)\right].
\end{split}
\end{equation}
where $q=(1-\mu )$, $a=(\epsilon+\sigma)$, and $b=(\epsilon+\sigma+\omega)$.  Unlike in case of other SEIRS models, in our model the probability of becoming infectious is not based on the mass action principle, but instead we use a biologically more sensible incidence function:  $G_t(i,j) = \rho (i, j) X_t(i,j)[1-\exp^{-\beta Y_t(i,j)}]$~\cite{jose}, where $\beta$ is a dimensionless constant representing the transmission parameter. 

The model exhibits a rich variety of behaviours, as expected: regular oscillations of different period, damped oscillations, quasi-periodicity, chaos and stable regimes. Therefore, the dynamics of other infectious diseases can be modelled with the same system, since all parameters so far, except the population density, characterise the disease, such as its virulence and lethality. In this study We shall give specific values to these parameters when applied to the specific country or region of COVID-19, as discussed below.

\subsection {Geographical disease spreading (Macrodynamics)}

In each cell the relative population is represented by a population density matrix $\rho(i,j)$. The size of each cell is determined by the actual conditions of the country or region to be simulated. The model in each cell represents an instantaneous local average state there, and the disease can be transmitted from one cell to another because people move distances larger than the cell size.
Therefore, the size of each cell should be congruent with the average distance that a person travels every day, to work or shopping, etc.
  
Here we consider three mobility mechanisms:
\begin{enumerate}
\item Cell to cell transmission is often modelled as a diffusion process, but this implies that the susceptible receives a certain amount of infection, and the infectious one becomes healthier by the same amount. Other SEIR models consider anomalous diffusion with fractional derivatives, which seems to be more appropriate \cite{chen}. An alternative approach is to consider a parameter of average terrestrial mobility or velocity, $v_t$, which we assume to be stochastic in nature. Therefore, we can model the probability of spreading the disease from one cell to neighbouring cells by using a Metropolis Monte-Carlo algorithm. First, one locates the potential spreader cells, this means having $Y_t(i,j) \ge \eta$, where $\eta$ is related to the infectiousness of the disease. Then, one chooses a random number from a uniform distribution ($p \in [0,1]$), and if $p$ is smaller than $v_t$, a neighbour cell (denoted by $\alpha$) becomes infected, that is $X(\alpha)=1-\eta$ and $Y(\alpha)=\eta$. It is clear that we know very little about the real meaning of $v_t$, since the reasons for people to making a trip are quite varied. In this sense, $v_t$ is related to the social and cultural habits of the individuals, and it is related to the average number of personal contacts per day.

\item Particular attention has been drawn on the influence of air travelling upon the spread of diseases~\cite{grais,grais1,longini,rvachev,flahault, brownstein, bekiros}. A similar stochastic process is used to model transmission from a cell to other more distant cells, connected by airlines. In this case, the probability of spreading the infection is proportional to the number of passengers per day travelling from one airport to another airport. This should be simulated by locating the airports in the grid and defining an air mobility parameter $v_a$. The probability of infection is represented by a weighted adjacency matrix ($A$) of the airline network, whose elements represent the average number of passengers between linked airports. Then, we run once again the Metropolis Monte-Carlo algorithm by using $v_a$ to decide infectiousness between the elements of $A$ (cells including the airports).

\item Since people seemingly travel randomly between distant places, noise has to be considered. This will cause the spread of the epidemics to unexpected places. In order to simulate this we introduce a Monte-Carlo procedure, similar to the geographical spread of the illness. Therefore, we consider cells $(i,j)$ in which there are not many susceptible ($X_t(i,j)<\eta$) at random, and compare the value of a random number $p_0$ with a quantity of the form $e^{-1/kT}$, since these random displacements can be considered analogous to the ``kinetic energy'' $kT$ or ``temperature'' $T$ of the system. If the Monte Carlo condition is fulfilled, then one starts the disease in that cell.
\end{enumerate}

\section{Application to the COVID-19 epidemics in Mexico, Finland and Iceland}

As in this model the biological parameters are clearly separated from those related to social, cultural, and economic phenomena, one could adjust the biological parameters according to present knowledge. For the COVID-19 the latency, $\epsilon$, is thought to be from 2 to 14~\cite{yang, rothe} days, though interestingly, $\epsilon=1$ improved the adjustment of the data; the infectiousness $\sigma=14$ is set to the standard “quarantine” time used in many countries; the immunity $\omega$ is yet unknown, but for SARS-CoV antibodies and memory T-cell response were detected 1 year or even longer after the infection  ~\cite{janice, ng}. To be conservative, here we use $\omega=140$.  Currently there are no data about transmission parameter $\beta$ and disease infectiousness parameters $\eta$, so we estimated them to be $\beta=0.91$ and $\eta=0.1$, by adjusting Mexican data from March 3 to April 13. As we assume that these parameters to be epidemiologically relevant for COVID-19, we use them unchanged for the three countries studied here. However, the mobility parameters were adjusted for each country taking into account different strategies and measures used in them to mitigate the effect of the pandemic. 

\subsection{The case of Mexico}

In order to illustrate the results obtained from the model, we use the geographical demographic data of Mexico from INEGI~\cite{inegi}  and from the John Hopkins University (J.H.) map~\cite{jh}  and apply it to the outbreak of COVID-19 in 2020. The frequency of air travel 
and the number of passengers per day in Mexico were obtained from Direcci\'on General de Aeron\'autica Civil of the Secretar\'{\i}a de Comunicaciones y Transportes~\cite{aero}.

In Mexico the first imported case occurred in Mexico City (central Mexico) on the 28th of February, and shortly after more imported cases were reported, one in Mexico City, other in Cancun (Yucatan peninsula) , other  in Baja California and another in Sinaloa (northern Mexico). Mexican government together with Ministry of Health (Secretaría de Salud) implemented some measures such as tracking COVID-19 positive people and their contacts. On March 14th, the Ministry of Public Education (Secretaría de Educación Pública) extended vacations in primary and high schools until April 20th in the whole country; some days latter this period was extended until April 30th. On March 18th university students were sent home and social distancing was initiated. On March 23th, at national level, Universities were closed as well as public places such as concerts, theatres, cinemas, churches, museums, gyms, zoos, and bars; home confinement was recommended. Airports were not closed, but they diminished their operations to only a small percentage of flights and passenger volumes. On March 26th non-essential activities of the government were suspended, and on March 30th non-essential economic activities at national level were suspended until April 30th. 

On April 13th we made a calculation adjusting the mobility parameters to the data available from the John Hopkins page~\cite{jh}. The values which best fitted the data were: $v_t=0.1$, $v_a=0.02$, and $kT=0.8$. These values represent an overall mobility reduction of approximately 35 \% of the normal values. There is also the possibility that the survival parameter  $S$ changes in time, if there are tests to detect potentially infectious people, or if there are vaccines or effective medication. So far this has not happened, so we used $S=0.1$, throughout the calculations.

On May 25th the Mexican authorities announced a plan to lift the restrictions gradually, and asynchronously, considering the situation of the pandemic in each state. Consequently, the three mobility parameters and the frequency of flights would change with time slowly, from May to August assuming that the mobility will be restored to 50\% of the value before the outbreak. 

 \begin{figure}[ht!]
\begin{center}
\includegraphics[width=.9\columnwidth]{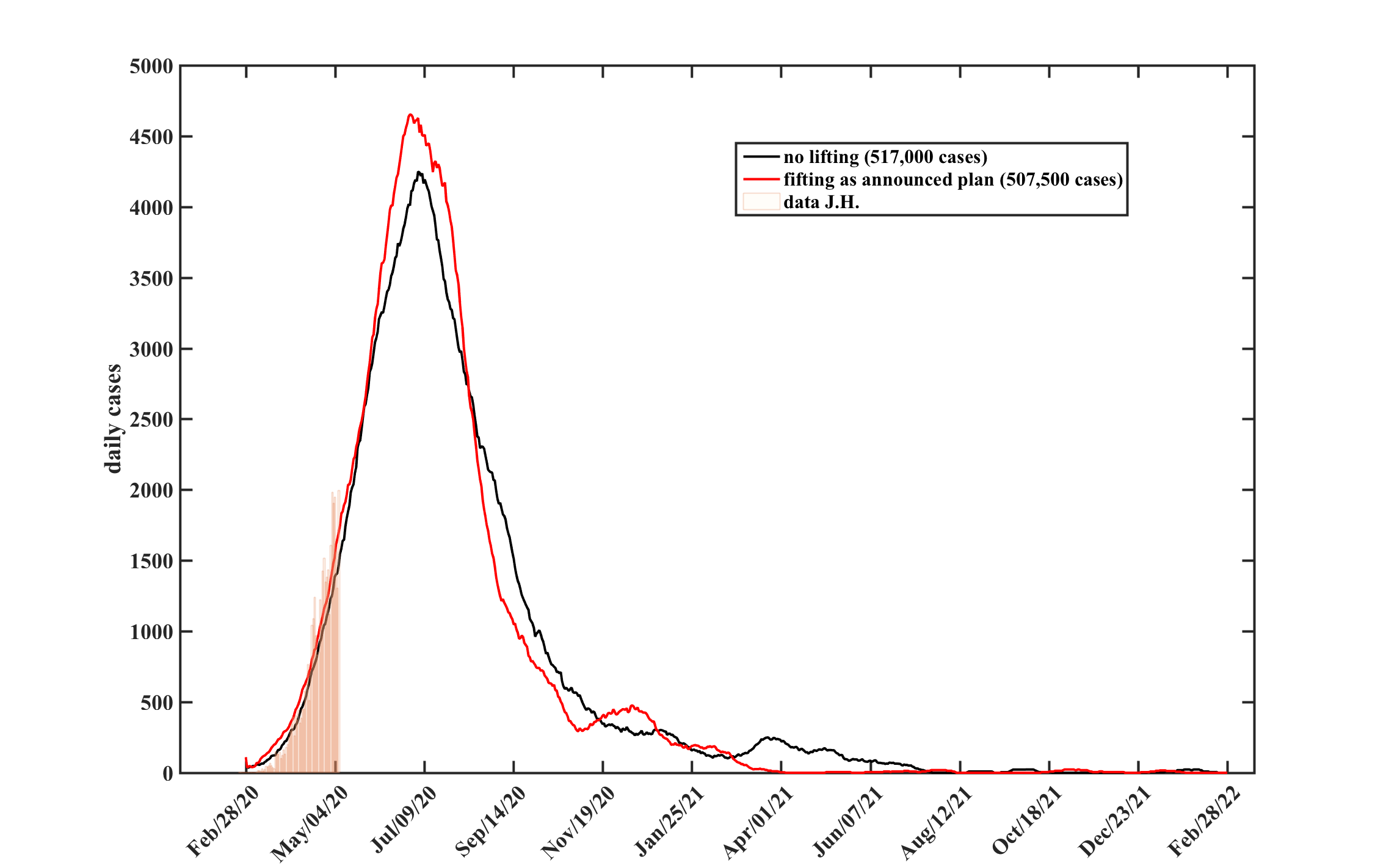} 
\caption{Geographical numerical calculations of the model with parameters appropriate for Covid-19. The prediction in this plot was made on April 13th, with the adjustments made from data up to that date. The black line  assumes that mobility restrictions do not change during the year, while the red line was obtained by feeding the mobility strategy announced by the government to lift restrictions gradually.}
\label{fig:1}
\end{center}
\end{figure}

In Fig.\ref{fig:1} we show examples of numerical calculations performed with these two hypothetical conditions, namely without lifting the restrictions, and following the plan to restore normality. Observe that the effect of lifting the restrictions is very small, but noticeable: the peak is attained around the 9th of July, in both cases, but taller when lifting, and the epidemic dies off faster than in the case of doing nothing. However, the number of infected people would remain practically the same.

\begin{figure}[ht!]
\begin{centering}
\includegraphics[width=.9\columnwidth]{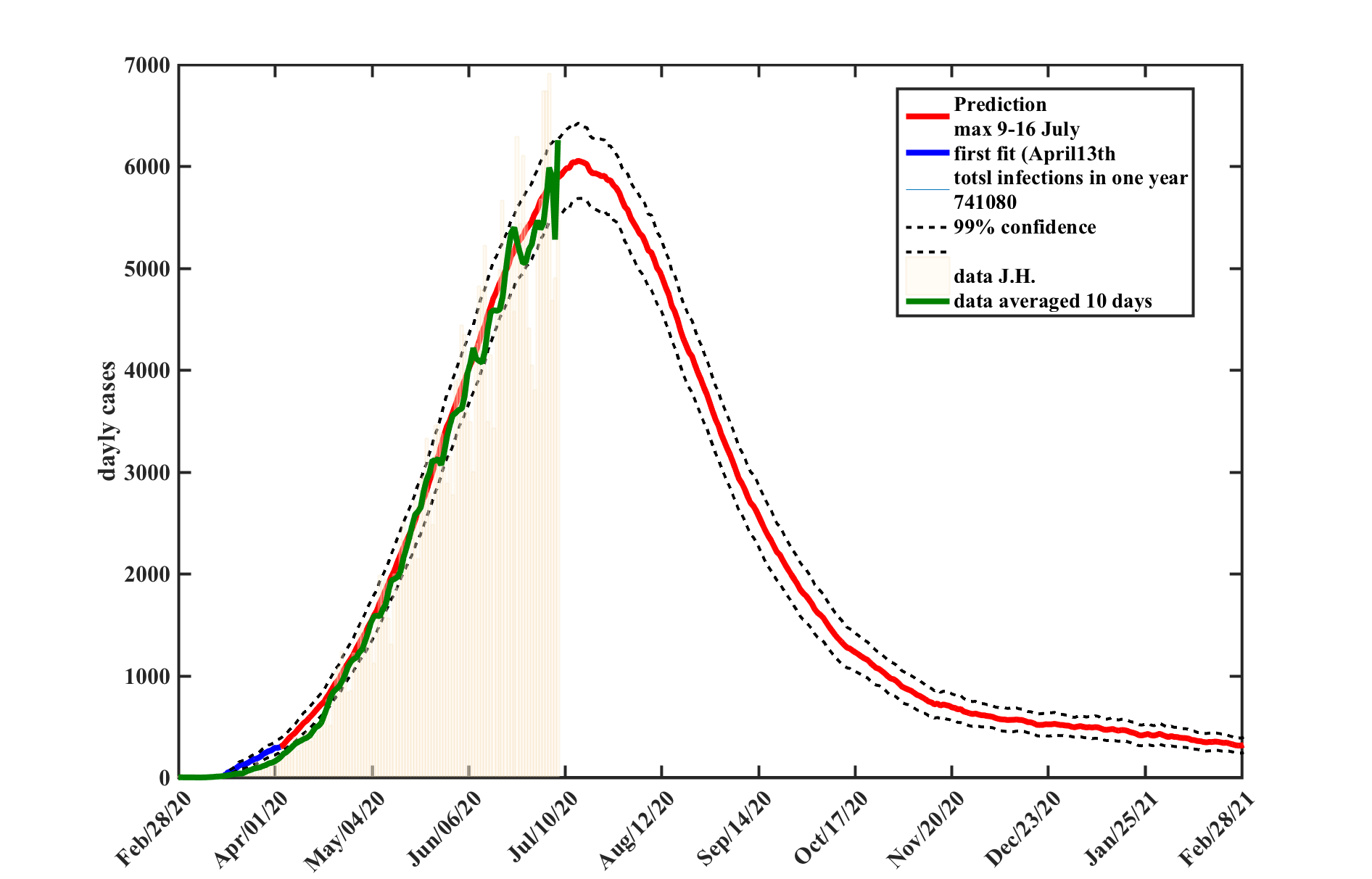}
\caption{Time history of the number of new cases per day. Continuous blue line:  daily confirmed  cases from February 28th to April 13th used to adjust the model parameters. Continuous red line:  numerical prediction assuming no further changes in societal conditions. Broken black lines : 99\% confidence interval. Bars: actual daily data up to July 8th. Green line is the 10-day average of the actual data.}
\label{fig:2}
\end{centering}
\end{figure}

In Fig.~\ref{fig:2} we show an average of 50 realisations of the time history of the pandemic and it is seen to model very closely the actual data. This means that although costly economically, the social distancing measures are extremely effective in mitigating the effects of epidemics. It is remarkable that the prediction made in April still holds tightly within the confidence interval more than two months later. A maximum of the number of cases is predicted to occur between the 9th and the 14th of July.

\subsection{The case of Finland}

On January 29th Finland confirmed its first COVID-19 case, a 32-year-old Chinese woman from Wuhan having sought medical attention in Ivalo and tested positive for SARS-CoV-2. She was quarantined at Lapland Central Hospital in Rovaniemi. The woman recovered and was discharged on February 5th after testing negative on two consecutive days. Three weeks later, on February 26th,  Finland's health officials confirmed the second case, a Finnish woman, who made a trip to Milan and returned on February 22th, tested positive at the Helsinki University Central Hospital. On February 28th, a Finnish woman who had also travelled to Northern Italy, tested positive by the Helsinki and Uusimaa Hospital District and was advised to remain in home isolation \cite{k1}. 

These two latter cases mark the beginning of COVID-19 epidemic in Finland, and the number of new cases was then followed closely by the National Institute of Health and Well being (THL) \cite{k2}, advising the Finnish Government the course of pandemic and the disease confinement measures \cite{k3}. As a result on March 12th the Government decides on recommendations to curb the spread of corona virus by public events and workplaces closures. This was soon followed on March 16th by  Government decision on lock down measures, to shut down all schools and most government-run public facilities (theatres, libraries, museums etc.). In addition, at most 10 people were allowed to participate in a public meeting, outsiders were forbidden from entering healthcare facilities and hospitals, and travel cross the internal Schengen and EU borders were limited to essential goods transportation, people having to because of work, and citizens returning to Finland. These measures were scheduled to be in place until April 13th, but were later extended to May 13th at which point the pre- and primary schools were opened for the last two weeks of the spring semester, while other measures were kept in effect.

On March 21st the first death, an elderly individual who lived in the Helsinki and Uusimaa hospital district, was reported. On March 27th, the Parliament voted unanimously to temporarily close the borders of the Uusimaa region (area surrounding the capital Helsinki), which had and still has the most confirmed cases. On April 15th travel restrictions between Uusimaa region and the rest of the country were lifted. In addition, from April 4th to May 31st the Government decided to close all restaurants and hotels. On June 1st onward the gradually opening of country started by allowing the maximum number of people to be increased to 50, opening restaurants and public places such as museums as well as allowing sports events, all of these "with special arrangements" to maintain sufficient physical or social distance. Then according to government decision from June 15th on, travellers entering from the Baltic countries and the other Nordic countries except Sweden will no longer have to stay quarantined for 14 days, but other international travel restrictions were kept. From July 1st onward Government decided that the outdoor events with more than 500 people will be allowed with the arrangements to keep sufficient physical distance. Overall it is worth mentioning that people have so far followed all the confinement measures, restriction and recommendations by the Government very closely, which is seen as a large reduction of reported daily infection cases to less than 10. 

Due to the above mentioned travel restrictions air travel has reduced to 2 - 3 \% from mid-March to May and to 6 - 7 \% in June in comparison to the same months last year (2019). The air travel volumes will very gradually start increasing from the beginning of July on, as the EU internal border restrictions are gradually started to be lifted. Due to strong recommendation by the Government to avoid travelling within Finland during mid-March to June period even the passenger volumes by railways have been running low, but expected to start picking up from mid-June on as people prefer having their vacations in Finland. We estimate the travel volumes only to about 20\% the value before the epidemic.
 
Taking these measures into account we introduced them in the model to modify the aerial, terrestrial and casual mobilities along the time line of their introduction. For the model there are four major events: (i) the 12-14 March closing down (with $v_t$, reduced to 50 \%, $ kT=0.4$), (ii) the March 27th isolation of the Uusimaa region (causing drastic reduction of $v_t$ to 25\% and $kT=0.1 $), (iii) the April 15th lifting travel restrictions of the Uusimaa region (with $v_t$ and $kT$ restored to the March values), and (iv) the air travel reduction from March to June, (with $v_a$ changing from 2\% in March - May to 7\% in June).

\begin{figure}[ht]
\begin{center}
\includegraphics[width=\columnwidth]{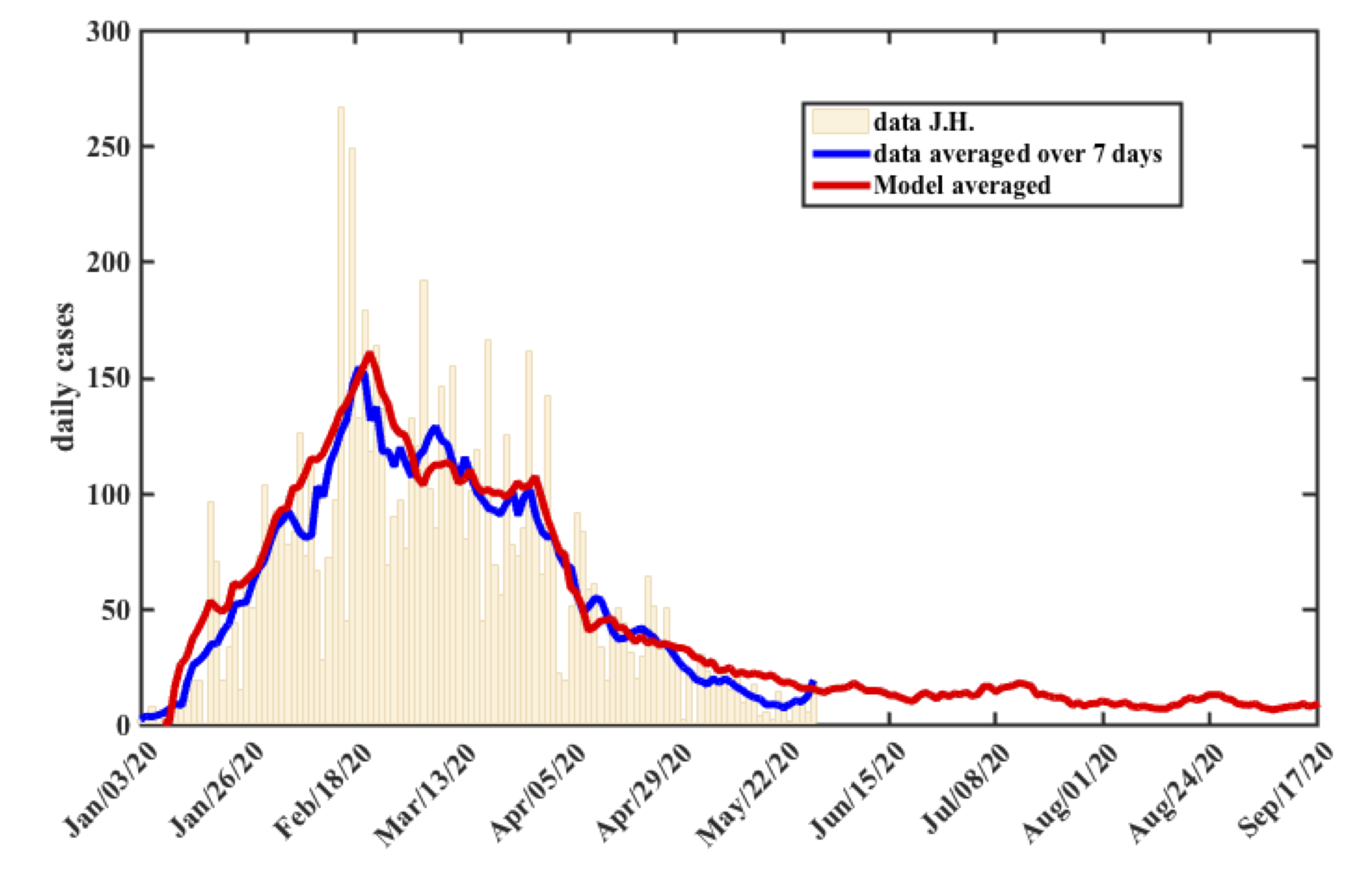} \\
\caption{Time history of the number of new cases per day in Finland. Continuous blue line: Actual data averaged over 7 days. Continuous red line:  numerical prediction including changes in social distancing. Bars: actual data taken from~\cite{jh}}
\label{fig:3}
\end{center}
\end{figure}

In Fig. \ref{fig:3} we show one example of the intricate calculation that resulted from the above described social distancing measures. Observe that the averaged values of daily cases follows remarkably well the averaged actual data.

\begin{figure}[ht!]
\begin{center}
\includegraphics[width=\columnwidth]{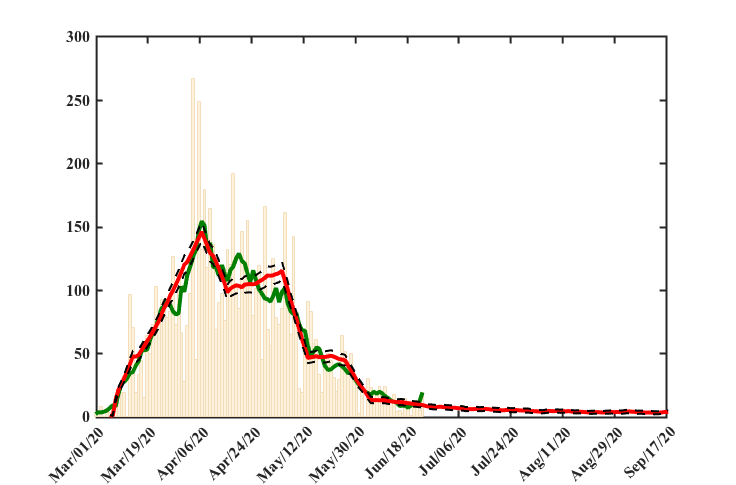} 
\caption{Average over 20 realisations of the model calculations (red line) showing the of 99\% confidence interval (black broken kines) and the averaged data in green. }
\label{fig:4}
\end{center}
\end{figure}

We have also performed a set of forty realisations and averaged them to calculate the of 99\% confidence interval of the model predictions. The results are shown in Fig. \ref{fig:4}. Observe that the model works quite well over the whole time interval, but little less so in between April 15th and May 5th, when the Uusimaa area opened. This is possibly due to people's heightened willingness after opening to go visit their relatives and their summer cottages. As we have no data of the number of people that travelled by road during that lapse, we have not been able to take it into account in the model.
\begin{figure}[ht!]
\begin{center}
\includegraphics[width=\columnwidth]{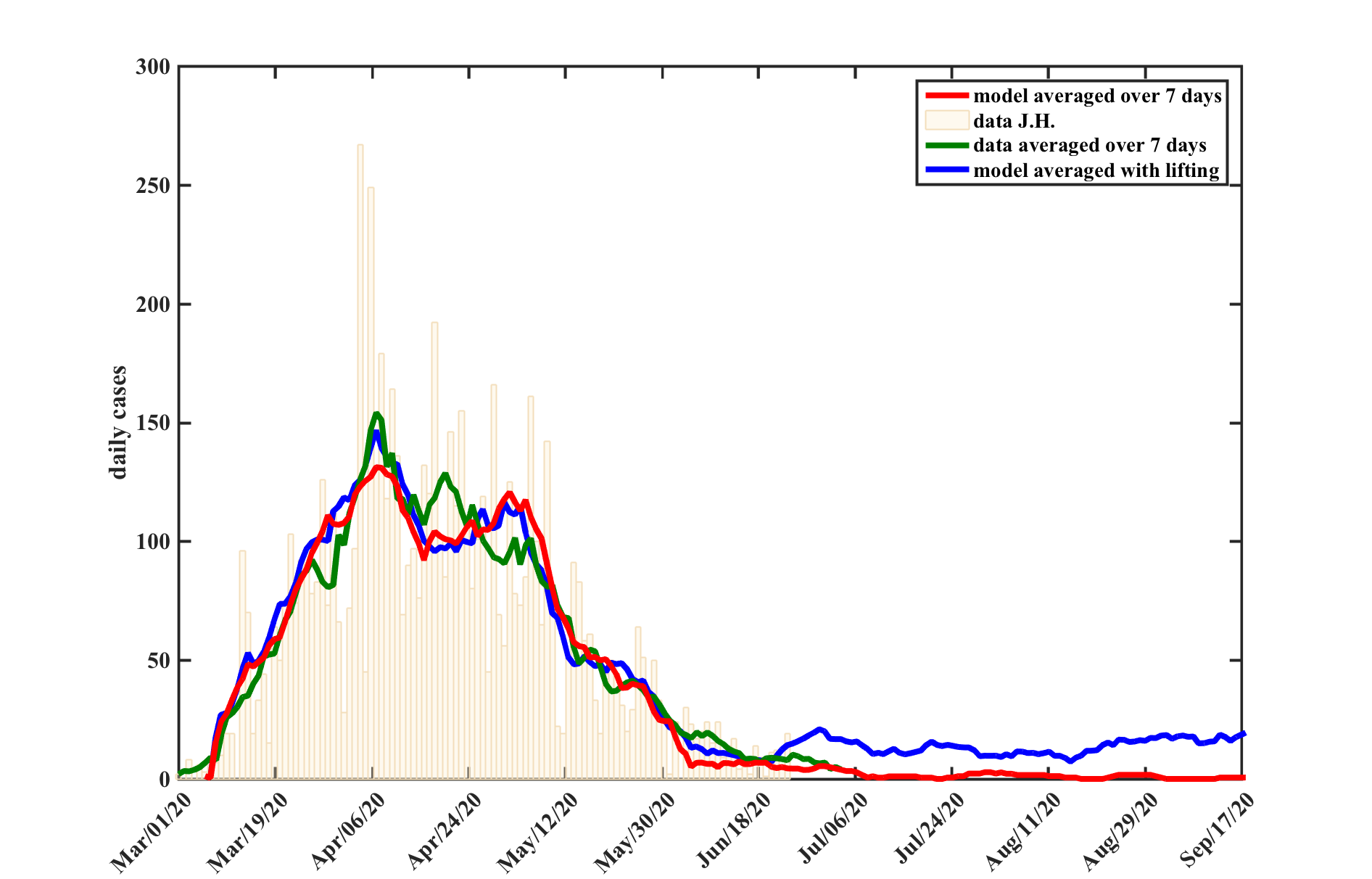} \\
\caption{As Fig. \ref{fig:4} including in blue the prediction if trains and airport start running on June 15th (trains  20\% and airport to 7\%)}
\label{fig:5}
\end{center}
\end{figure}
 Just as a matter of curiosity we made a calculation to predict the effect of train travel passenger volumes starting from June 15th to increase gradually up to 20\% and air travel passenger volumes gradually up to 7\% of the values before the pandemic. The results are depicted in Fig. \ref{fig:5} and show only a slight increase in the number of daily cases to about a dozen and persisting at that level throughout summer.  
\subsection{The case of Iceland}

On January 29th, Chief Epidemiologist Thorolfur Gu\dj nason advised against unnecessary travel to China, and recommended that people travelling from China undertake 14 days quarantine upon returning to Iceland.  On the 31st of January, a meeting in the National Security Council was scheduled with the Minister of Health and Chief Epidemiologist. The Department of Civil Protection and Emergency Management (DCPEM) evoked the National Crisis Co-ordination Center. By the 3rd of February, Iceland had defined high-risk areas, including Northern Italy and Tirol, earlier than other governments, taking stricter measures with a 14-day quarantine requirement for all residents returning from those areas.

February 27th saw the first daily press conference, attended by Iceland’s Chief Epidemiologist, Director of Health and Chief Superintendent. The first case of COVID-19 was confirmed in Iceland on February 28th. This was a person arriving from Northern Italy. DCPEM declared the alert phase. By March 6th  over 30 imported cases had been confirmed and the first two transmissions within Iceland, traced to infected individuals who had recently traveled to Northern Italy. The alert level was raised to the emergency phase.

On March 13th, screening for the virus that causes COVID-19 started among the general public. deCODE genetics graciously offered to test everyone in the country who wanted to be tested. At the same time, a ban on gatherings of more than 100 people was announced and then implemented on the 16th. High schools and Universities were closed, and operations of kindergartens and primary schools were limited. 

By March 31st, Iceland had limited gatherings to 20 people or fewer, closed sports clubs, hair salons, bars, and similar establishments, implemented fines for breaching the rules of quarantine, and became a party to an international contract which enabled the Icelandic authorities to join European Union members in the procurement of various healthcare equipment. April 2nd saw the tracing app Rakning C-19 become available in the App Store and Google Play to track the virus. 

As a result of the quick action taken by the government, results from initial screenings indicated a low rate of infections among the general public. By April 21st, it was announced that the ban on gatherings and school activities would be relaxed, effective on May 4th 2020. The limit on the number of people who may gather increased from 20 to 50; pre-schools and primary schools reopened; athletic and youth activities became unrestricted again.

On May 25th, gyms and swimming pools reopened and operated with limitations, such as a maximum number of guests.  Gatherings of up to 200 people were allowed. All restaurants and bars re-opened with a curfew of 11 PM \cite{ice_dh_2020}.

Taking these measures into account in the model we changed only the mobility parameters on the dates mentioned. We considered domestic flights operating with very few and diminishing number of passengers from the beginning of the outbreak, while the international airport was getting many imported cases until March 16th, when the aerial and terrestrial mobility were assumed to be reduced by 70\%, reflecting the full obedience of the population to the confinement measures taken. Between April 2nd and April 9th a 90\% mobility reduction was achieved, but on May 5th the mobility increased to 60\%. The random mobility was kept to the original value of $kT=0.8$, because of the uneven population density in the island, as 62\% of the population is concentrated around the Reykjavik area, so it makes little difference if $kT$ is reduced.

\begin{figure}[ht!]
\begin{center}
\includegraphics[width=\columnwidth]{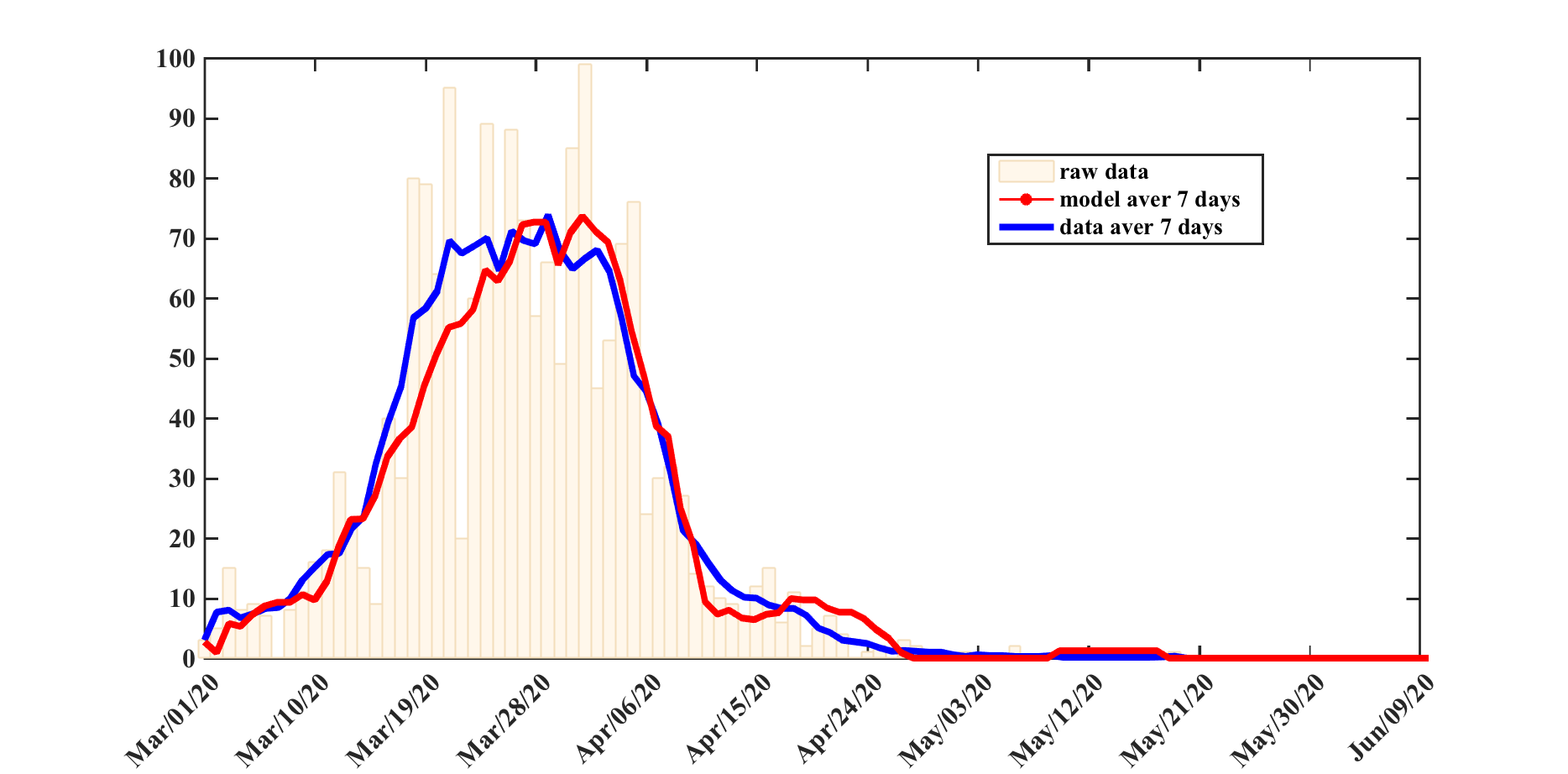}
\caption{(a) Time history of the number of new cases per day in Iceland. Continuous blue line: Actual data averaged over 7 days. Continuous red line:  numerical prediction including the changes in social distancing mentioned in the text. Bars: actual data taken from~\cite{jh}}
\label{fig:6}
\end{center}
\end{figure}

In Fig.\ref{fig:6} we show one realisation of the model and compare it with the actual data, both averaged over 7 days. We could say that in this case, the disease had been controlled in the early days of May. 

The borders opened on June 15th. Passengers arriving in the country were given the option of taking a COVID-19 test or undergoing quarantine for 14 days. Children born in 2005 and later were exempt. Testing was offered at Keflavík Airport and at other international ports of entry. Passengers were also required to answer a pre-arrival questionnaire, abide by sickness rules and were encouraged to download the app, Rakning C-19. At the same time, further relaxation of the ban on assembly took effect. The number of gatherings increased from 200 to 500 and restrictions on the number of swimming pools and fitness centers were reduced.

\begin{figure}[ht!]
\begin{center}
\includegraphics[width=\columnwidth]{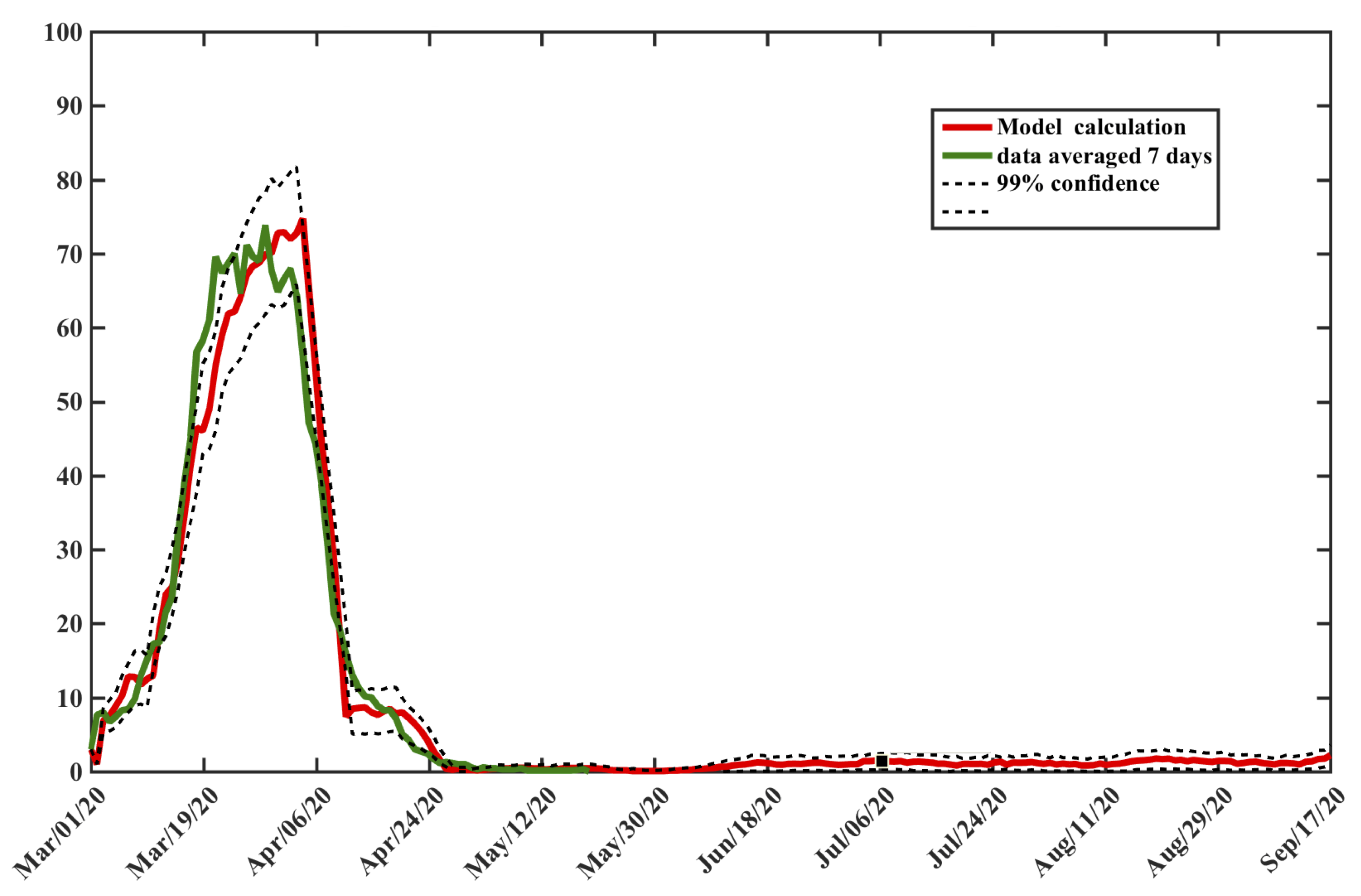}
\caption{(a) Average over 20 realisations of the time history of the number of daily cases in Iceland. Continuous green line: Actual data averaged over 7 days. Continuous red line:  numerical prediction including the opening of the airport in Keflavik on June 15th. Broken lines: 99\% confidence interval.}
\label{fig:7}
\end{center}
\end{figure}

Considering these relaxations of the social distancing measures, we performed a calculation on June 10th to try predicting the effects that would result from this action. In Fig.\ref{fig:7} we show the results from the model averaging over 20 realisations. We assume that one should expect around 1.5 cases per 1000 passenger coming from abroad.

The results are encouraging, since the model is predicting a tail of 2 to 3 new daily cases on average. Observe that the confidence intervals are larger than in the cases of Finland and Mexico. This is expected, since the population, and consequently, the number of cases is much smaller. Despite the small number of cases predicted when opening the airport, people in Iceland received a warning, they were too relaxed in early June, particularly young people thinking that the pandemic was over in June, but after some new cases appeared after June 15th, they became more careful, scared and worried and started behaving accordingly.

\section{Discussion}

We have presented a stochastic model of geographical spreading of infectious disease that differs in many respects from other compartmental SEIRS type models, commonly used to describe epidemic spreading. In particular, our model separates the disease-defining epidemiological parameters from the ones that have to do with geography, and people's social habits. It also includes societal traffic and travelling infrastructure and passenger flow networks based on roads, trains and airline routes. Furthermore, it allows one to define mobility quantities that could vary in time, to make predictions of future scenarios of the epidemic.

There is a recent review \citep{vesp} that discusses various modelling approaches to study particularly the COVID-19 pandemic, and the general comment is that for models to be more predictive, one needs much more and precise data. The fact that with the model presented here one can make quite accurate predictions is due to its intrinsic random nature, which allows to approach real situations with very little information. This model has performed well also with other viral diseases, as shown earlier with the swine influenza \cite{barrio}. With the present study we have demonstrated that this model can be used for very different scales of population and their densities as well as for countries with different social structures and situations. 

By studying the three countries presented here we are able to learn several important lessons. First of all that it could be unwise to lift the country or regional level confinement measures, restrictions and recommendations completely when the pandemic is still active in other countries or regions, as it was demonstrated in the case of Iceland and Finland.  
It shows that going back to a normality as free as last year will not be possible for a long time. Of course we hope that there is an effective vaccine or treatment available soon. It should be noted that in the case of Mexico the model prediction of the behaviour of the pandemic was made early in the so called exponential phase, i.e. April 13th when the amount of available information was small, yet it has followed the actual number of cases very closely so far. 

\begin{figure}[ht!]
\begin{center}
\includegraphics[width=.85\columnwidth]{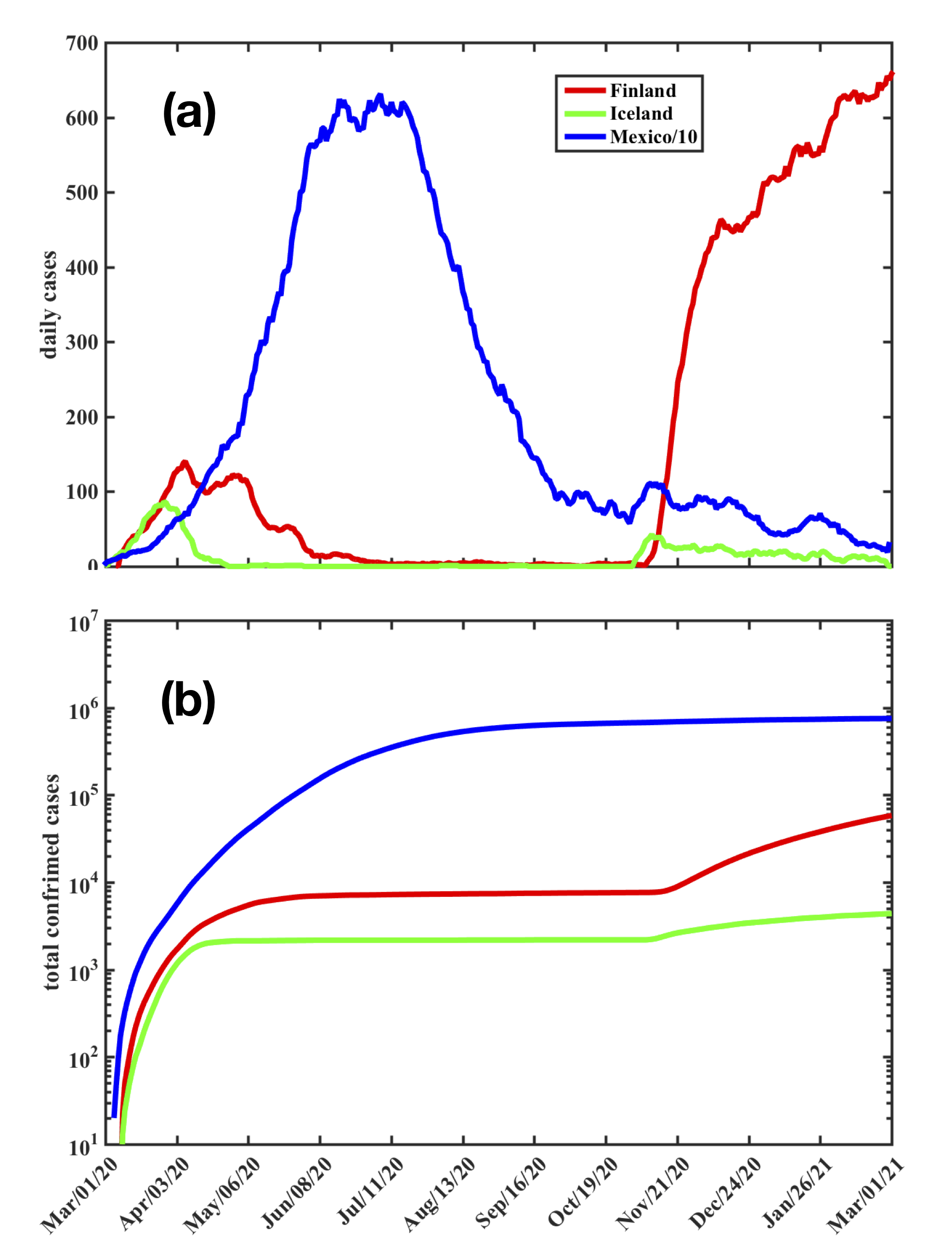}
\caption{(a) Calculations comparing the number of daily cases for the three countries in the hypothetical scenario that from November 6th onward the mobility is reestablished as it was in December 2019. Notice that the data for Mexico have been divided by 10. (b) Same comparison for the total number of infections in logarithmic scale.}
\label{fig:8}
\end{center}
\end{figure}

The second issue is that of the so called second wave, to be expected around the end of this year or beginning of next year. So far we have made the calculations assuming that the countries remain very much isolated, with very limited travel between them. In addition we have used this model to foresee what could happen if the three countries were made totally open by the end of this year. The results are shown in Fig.\ref{fig:8}. 

We observe that the rebound is greatest in Finland, small but noticeable in Iceland and extremely small in Mexico. This is understandable on the grounds that in Finland a large portion of the population has not become exposed during the first wave due to obediently followed confinement measures, in Iceland this is not the case, due to obediently followed confinement measures and perhaps the most vivid population-wide testing and tracking policy 
and in Mexico the pandemics following a normal course infecting practically everybody and reaching even the most remote areas of the country during the first wave. 

Perhaps one of the most important points of this comparative study is that the timely social distancing and confinement measures, as well as public restrictions and recommendations are seemingly efficient ways to control the disease spreading, especially when people follow them tightly, as it is the case in Finland and Iceland both being sparsely populated and socially relatively homogeneous countries. In addition, in case of Iceland from the very beginning of the epidemic the population-wide testing and tracking served as an enhanced control and life-saver. In case of Mexico imposing confinement measures is more complicated, because of much larger population and very large densely populated cities with huge socioeconomic differences. Still the authorities have managed to slow down the speed of disease spreading. The price to pay in all these three countries is that the pandemic will be active for quite a while, but on the other hand our understanding of confinement and prevention measures has been tested and will allow to face the possible second wave in a better way.

\section*{Acknowledgments}
RAB acknowledges support from The National Autonomous University of Mexico (UNAM) and Alianza UCMX of the University of California (UC), through the project included in the Special Call for Binational Collaborative Projects addressing COVID-19. RAB was financially supported by Conacyt through project 283279. KK acknowledges support for Visiting Fellowship at The Alan Turing Institute, UK, and the European Community’s H2020 Research Infrastructures "SoBigData++: Social Mining and Big Data Ecosystem” project.

\bibliographystyle{unsrt}
\bibliography{influenza1}

\begin{thebibliography}{10}

\bibitem{and}
R.~M. Anderson and R.~M. May.
\newblock {\em Infectious diseases of human. Dynamics and control}.
\newblock Oxford Univ. Press, Oxford, 1992.

\bibitem{bailey}
N.T.J. Bailey.
\newblock {\em The Mathematical Theory of Epidemics.}
\newblock Charles Griffin, London, 1957.

\bibitem{cast}
Castillo-Ch\'avez and A.~Yakubo.
\newblock Discrete time sis models with simple and complex dynamics.
\newblock In C.~Castillo-Ch{\'a}vez, S.~Blower, P.~van~den Driessche,
  D.~Kirshner, and A.~Yakubo, editors, {\em Mathematical approaches for
  emerging and reemerging infectious diseases. An introduction}, pages
  153--163. Springer-Verlag, 2002.

\bibitem{johansen}
A.~Johansen.
\newblock A simple model of recurrent epidemics.
\newblock {\em J. Theor. Biol.}, 178:45--51, 1996.

\bibitem{london}
W.~P. London and J.~A. Yorke.
\newblock Recurrent outbreaks of measles, chickenpox and mumps: I. seasonal
  variation in contact rates.
\newblock {\em Am. J. Epidemiol.}, 98:453--468, 1973.

\bibitem{bootsma}
M.C.J. Bootsma and N.M. Ferguson.
\newblock The effect of public helth measures on the 1918 influenza pandemic in
  u.s. cities.
\newblock {\em Proc. Natl. Acad. Sci. USA}, 104:7588--7593, 2007.

\bibitem{viboud}
C.~Viboud, P.~Y. Bo\"elle, K.~Pakdaman, F.~Carrat, A.~J. Valleron, and
  A.~Flahault.
\newblock Influenza epidemics in the united states, france, and australia,
  1972--1997.
\newblock {\em Emerg. Infect. Dis.}, 10:32--39, 2004.

\bibitem{longini05}
Ira M.~Jr. Longini, Azhar Nizam, Shufu Xu, Kumnuan Ungchusak, Wanna
  Hanshaoworakul, Derek A.~T. Cummings, and M.~Elizabeth Halloran.
\newblock Containing pandemic influenza at the source.
\newblock {\em Science}, 309:1083--1087, 2005.

\bibitem{ferguson}
S.~Cauchemez, A.J. Valleron, P.Y. Bo\"elle, A.~Flahault, and N.M. Ferguson.
\newblock Estimating the impact of school closure on influenza transmission
  from sentinel data.
\newblock {\em Nature}, 452:750--755, 2008.

\bibitem{Balcan}
Duygu Balcan, Vittoria Colizza, Bruno Gon\c{c}alves, Hao Hud, Jos\'e~J.
  Ramascob, and Alessandro Vespignani.
\newblock Multiscale mobility networks and the spatial spreading of infectious
  diseases.
\newblock {\em Proc. Natl. Acad. Sci. USA}, 106(51):21484--21489, 2009.

\bibitem{barabasi}
F.~Simini, M.C. Gonz\'alez, A.~Maritan, and Barab\'asi A.L.
\newblock A universal model for mobility and migration patterns.
\newblock {\em Nature}, 484:96--100, 2012.

\bibitem{barrio}
R.A Barrio, C.~Varea, T.~Govezensky, and M.V. Jos\'e.
\newblock Modeling the geographical spread of influenza a(h1n1): The case of
  mexico.
\newblock {\em Applied Mathematical Sciences}, 7(44):2143--2176, 2013.

\bibitem{apolloni}
A.~Apolloni, C.~Poletto, J.J. Ramasco, P.~Jensen, and Colizza V.
\newblock Metapopulation epidemic models with heterogeneous mixing and travel
  behaviour.
\newblock {\em Theor Biol Med Model}, 1:3:PubMed: 24418011, 2014.

\bibitem{marguta}
R.~Marguta and A.~Parisi.
\newblock Impact of human mobility on the periodicities and mechanisms
  underlying measles dynamics.
\newblock {\em J. R. Soc. Interface}, 12:PubMed: 25673302, 2015.

\bibitem{santermans}
E.~Santermans, E.~Robesyn, T.~Ganyani, B.~Sudre, C.~Faes, C.~Quinten,
  W.~Van~Bortel, T.~Haber, T.~Haber, F.~Van~Reeth, M.~Testa, N.~Hens, and
  D.~Plachouras.
\newblock Spatiotemporal evolution of ebola virus disease at sub-national level
  during the 2014 west africa epidemic: Model scrutiny and data meagreness.
\newblock {\em PLoS One}, 11(1):e0147172, 2016.

\bibitem{chowell}
G.~Chowell, L.~Sattenspiel, S.~Bansal, and C.~Viboud.
\newblock Mathematical models to characterize early epidemic growth: A review.
\newblock {\em Phys Life Rev.}, 18:66--97, 2016.

\bibitem{chen}
Xu~Conghui, Yu~Yongguang, Chen YangQuan, and Lu~Zhenzhen.
\newblock Forecast analysis of the epidemics trend of covid-19 in the united
  states by a generalized fractional-order seir model.
\newblock {\em Nonlinear Dyn (to be published)}, 2020.

\bibitem{chen2}
Lu~Zhenzhen, Yu~Yongguang, Chen YangQuan, Ren Guojian, Xu~Conghui, Wang Shuhui,
  and Yin Zhe.
\newblock A fractional-order seihdr model for covid-19 with inter-city
  networked coupling effects.
\newblock {\em arXiv}, page 12308v3, 2020.

\bibitem{yang}
Yang Chayu and Wang Jin.
\newblock A mathematical model for the novel coronavirus epidemic in wuhan,
  china.
\newblock {\em Mathematical Biosciences and Engineering}, 17(3):2708--2724,
  2020.

\bibitem{bekiros}
S.~Bekiros and D.~Kouloumpou.
\newblock A new mathematical model of infectious disease dynamics.
\newblock {\em Chaos, Solitons and Fractals}, 136:109828, 2020.

\bibitem{jose}
M.V. Jos\'e, T.~Govezensky, A.V. Lara~Sagahon, C.~Varea, and R.A Barrio.
\newblock A discrete seirs model for pandemic periodic infectious diseases.
\newblock {\em Advanced Studies in Biology}, 4:153--174, 2012.

\bibitem{grais}
R.~F. Grais, J.~H. Ellis, A.~Kress, and G.~E. Glass.
\newblock Modeling the spread of annual influenza epidemics in the u.s.: The
  potential role of air travel.
\newblock {\em Health Care Manag. Sci.}, 7:127--134, 2004.

\bibitem{grais1}
J.~H. Grais, R. F .and~Ellis and G.~E. Glass.
\newblock Assessing the impact of airline travel on the geographic spread of
  pandemic influenza.
\newblock {\em Eur. J. Epidemiol.}, 18:1065--1072, 2003.

\bibitem{longini}
I.~M.~Jr. Longini, P.~E. Fine, and S.~B. Thacker.
\newblock Predicting the global spread of new infectious agents.
\newblock {\em Am. J. Epidemiol.}, 123:383--391, 1986.

\bibitem{rvachev}
L.~Rvachev and I.~M.~Jr. Longini.
\newblock A mathematical model for the global spread of influenza.
\newblock {\em Math. Biosci.}, 75:3--22, 1985.

\bibitem{flahault}
A.~Flahault, S.~Deguen, and A.~J. Valleron.
\newblock A mathematical model for the european spread of influenza.
\newblock {\em Eur. J. Epidemiol.}, 10:471--474, 1994.

\bibitem{brownstein}
J.~S. Brownstein, C.~J Wolfe, and K.~D. Mand.
\newblock Empirical evidence for the effect of airline travel on inter-regional
  influenza spread in the united states.
\newblock {\em PLoS Medicine}, 3(10):e401, 2009.

\bibitem{rothe}
C.~Rothe, M.~Schunk, P.~Sothmann, G.~Bretzel, G.~Froeschl, C.~Wallrauch, and
  al. et.
\newblock Transmission of 2019-ncov infection from an asymptomatic contact in
  germany.
\newblock {\em N. Engl. J. Med.}, 2020.

\bibitem{janice}
Oh.H.L. Janice, A.~Ken-En, G.S. end~Bertoletti, and Y.J. Tan.
\newblock Understanding the t cell immune response in sars coronavirus
  infection.
\newblock {\em Emerg Microbes Infect}, 1(9):e23, 2012.

\bibitem{ng}
O.W. Ng, A.~Chia, A.T. Tan, R.S. Jadi, H.N. Leong, A.~Bertoletti, and Y.J. Tan.
\newblock Responses targeting the sars coronavirus persist up to 11 years
  post-infection.
\newblock {\em Vaccine}, 34(17):2008--14, 2016.

\bibitem{inegi}
Instituto~Nacional de~Estad{\'\i}stica~y Geograf{\'\i}a.
\newblock https://www.inegi.org.mx, 2020.

\bibitem{jh}
John Hopkins University~(J.H.) map.
\newblock https://coronavirus.jhu.edu/map.htm, 2020.

\bibitem{aero}
Direcci{\'o}n~General de~Aeron{\'a}utica Civil of the Secretar{\'\i}a de
  Comunicaciones~y Transportes.
\newblock https://www.sct.gob.mx, 2020.

\bibitem{k1}
Coronavirus disease 2019.
\newblock https://en.wikipedia.org/wiki/covid-19.

\bibitem{k2}
Finish~Institute of~Health and Wellfare.
\newblock https://thl.fi/en/web/thlfi-en,.

\bibitem{k3}
Information and advice on~the coronavirus.
\newblock https://valtioneuvosto.fi/en/information-on-coronavirus.

\bibitem{ice_dh_2020}
The~Directorate of~Health, The~Department of~Civil~Protection, and Emergency
  Management.
\newblock https://www.covid.is/sub-categories/icelands-response, 2020.

\bibitem{vesp}
Alessandro Vespignani, Huaiyu Tian, Christopher Dye, James~O. Lloyd-Smith,
  Rosalind~M. Eggo, Munik Shrestha, Samuel~V. Scarpino, Bernardo Gutierrez,
  Moritz U.~G. Kraemer, Joseph Wu, Kathy Leung, and Gabriel~M. Leung.
\newblock Modelling covid-19.
\newblock {\em Nature Reviews}, 2:279--281, 2020.

\end{thebibliography}

\end{document}